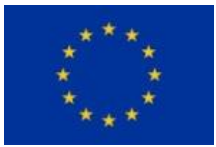
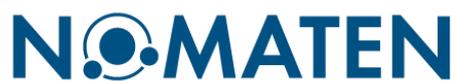

This work was carried out in whole or in part within the framework of the NOMATEN Centre of Excellence, supported from the European Union Horizon 2020 research and innovation program (Grant Agreement No. 857470) and from the European Regional Development Fund via the Foundation for Polish Science International Research Agenda PLUS program (Grant No. MAB PLUS/2018/8), and the Ministry of Science and Higher Education's initiative "Support for the Activities of Centers of Excellence Established in Poland under the Horizon 2020 Program" (agreement no. MEiN/2023/DIR/3795).

The version of record of this article, first published in Chemical Engineering Journal, Volume 518, 15 August 2025, 164510, is available online at Publisher's website:
https://doi.org/10.1016/j.cej.2025.164510

# Synergistic modulation of band structure and phonon transport for higher thermoelectric performance of WSe2

*Mazhar Hussain Danish [a,b,c], Amil Aligayev [a,b,e], Zahir Muhammad [d], Tao Chen [a,b], Adil Mansoor [c], Zia Ur Rahman [c], F. J. Dominguez-Gutierrez [e], Di Li [a], Jian Zhang [a\*], Zhuang-Hao Zheng [c\*], Xiaoying Qin [a\*]*

[a]Key Lab of Photovoltaic and Energy Conservation Materials, Institute of Solid State Physics, HFIPS, Chinese Academy of Sciences, Hefei 230031, P. R. China

[b]University of Science and Technology of China, Hefei, 230026, China.

[c]Shenzhen Key Laboratory of Advanced Thin Films and Applications, Key Laboratory of Optoelectronic Devices and Systems of Ministry of Education and Guangdong Province, College of Physics and Optoelectronic Engineering, Shenzhen University, Shenzhen, Guangdong, 518060 P. R. China

[d]Hefei Innovation Research Institute of Beihang University, Hefei, 230013, China

[e]NOMATEN Centre of Excellence, National Center for Nuclear Research, 05-400 Swierk/Otwock, Poland.

**KEYWORDS**: thermoelectric materials; thermopower; co-doping; phonon scattering; point defect; mobility;

\*Corresponding Author

E-mail address:    zhangjian@issp.ac.cn      (J. Zhang)
                   zhengzh@szu.edu.cn        (Z.H. Zheng)
                   xyqin@issp.ac.cn          (X.Y. Qin)

**ABSTRACT:**

Tungsten diselenide (WSe$_2$) emerges as a promising thermoelectric (TE) candidate due to its high thermopower (*S*), cost-effectiveness, and environmentally friendly characteristics. However, the pristine WSe$_2$ exhibits limited electrical conductivity ($\sigma$), a low power factor (*PF*), and high lattice thermal conductivity ($\kappa_L$), which restrict its overall TE performance. Here, we show that via co-doping of Nb for W and Te for Se in WSe$_2$, its power factor (*PF*) undergoes 17-fold increase, reaching 8.91 µWcm$^{-1}$K$^{-2}$ at 850 K. Simultaneously, its lattice thermal conductivity ($\kappa_L$) decreases from 1.70 Wm$^{-1}$K$^{-1}$ to 0.48 Wm$^{-1}$K$^{-1}$. Experiments and DFT analysis demonstrate that the enhancement of the *PF* is linked to enhanced density of state, effective mass ($m_d^*$), improved mobility ($\mu$) and elevated electrical conductivity ($\sigma$) owing to replacing Se$^{2-}$ with Te$^{2-}$; while the observed 72% reduction in $\kappa_L$ results primarily from phonon scattering at defects Te$_{Se}$ and Nb$_W$. As a result, a remarkable *ZT$_{max}$* ~ 1 is obtained at 850 K for the sample W$_{0.95}$Nb$_{0.05}$Se$_{2-y}$Te$_y$ with *y*=0.3, which is ~ 30-fold increase than that of WSe$_2$, proving that Nb and Te co-doping in WSe$_2$ can significantly boost its TE performance.

# 1. Introduction:

In response to the increasing energy demands, there is a growing potential on the development of novel energy sources. Among these, TE materials have emerged as a promising solution for efficiently transforming waste heat into electrical energy, offering both environmental and economic benefits.[1–3] TE materials have gained interest recently due to their distinct features, which are frequently utilized in the military, aerospace, medical, communications, wearable technology, and other industries due to its compact size, high reliability, and broad temperature range.[4–10] Currently, the thermoelectric figure of merit ($ZT=\sigma S^2 T/\kappa$) value remains a limit on the conversion efficiency of thermoelectric materials, where $\kappa$, $S$, $\sigma$, and $T$ stand for the total thermal conductivity, thermopower, electrical conductivity, and absolute temperature, respectively; Whereas, total thermal conductivity ($\kappa$) is defined as ($\kappa = \kappa_e + \kappa_L$), while $\kappa_e$ represents electronic thermal conductivity and $\kappa_L$ represents lattice thermal conductivity.[11–14] The coupling between TE transport parameters makes it very difficult to obtain high $ZT$ values. However, one must simultaneously reduce thermal conductivity and increase power factor ($PF=\sigma S^2$) in order to obtain high $ZT$.[15] Two-dimensional TE materials $PF$ can be significantly increased by enhanced value of $\sigma$ and $S$.[16] The numerous interfacial scattering between layers can increase phonon scattering, resulting in lower lattice thermal conductivity.[17–19] Consequently, two-dimensional materials exhibit significant promise as TE candidates, thus attracting substantial interest in investigating their thermoelectric properties within the scientific community.[20,21] Although transition metal dichalcogenides are typical two-dimensional thermoelectric materials, their $ZT$ value is not ideal due to significant coupling among TE factors ($\sigma$, $\kappa$, and $S$) by applying simple approach.[22,23]

Transition-metal dichalcogenides ($ZM_2$, where Z represents a transition metal and M chalcogenide atom) belong to layered materials with their component elements being comparatively cheap as well as sustainable in nature. It is a compound where the layers of Z and M have strong covalent bonds and weak Van der Waals forces.[24,25] They have attracted a lot of interest since they offer a wide range of application potentials, and including semiconducting, superconducting, lubricating, and magnetic features. Transition metal disulfides have been the focus of extensive theoretical as well as experimental studies over the last few decades ($WS_2$, $MoS_2$, and $TiS_2$) as TE materials.[26–31] The TE performance of transition-metal disulfide is often relatively limited due to its high thermal conductivity (for example, higher than that of 5 W m$^{-1}$ K$^{-1}$). In contrast, within the hexagonal family, $WSe_2$ exhibits a high atomic weight but a low thermal conductivity. According to published research on WSe$_2$ thin films, its thermal conductivity is substantially lower (0.5 W m$^{-1}$ K$^{-1}$) than transition-metal disulfide (MoS$_2$, WS$_2$, or TiS$_2$). Moreover, intrinsic WSe$_2$ has very high $S$ values and exhibits p-type semiconductor behavior. As a consequence, WSe$_2$ might be a promising challenger for good TE applications.[32–41] In order to compare TE performance in perpendicular but also parallel directions to the direction Liu et al. determined the TE efficiency of WSe$_2$ at different spark plasma sintering (SPS) temperature range (*ZT*, 0.03 at 900 K).[42] In a W$_{1-x}$Ta$_x$Se$_{1.6}$Te$_{0.4}$ composite, Kriener et al. discovered a relatively large PF value and $ZT_{max}$ ~ 0.35 at elevated temperature 850 K.[43] By substituting Nb at a room temperature (300 K), Yakovleva et al. were able to determine the thermoelectric efficiency of WSe$_2$ (*ZT*, 0.02).[44] Prior to this, they have also explored the co-doping of Nb on W and the partial replacement of S on Se and achieved *ZT* 0.26 at 650 K.[45] Danish et al. conducted an experimental study to determine the thermoelectric performances of Nb-doped WSe$_2$ at high temperatures (between 300 K to 850 K) and obtained moderate value of *ZT* ~ 0.42.[46] Danish et al. recently extended their experimental research to assess the thermoelectric performance of the Nb-doped quarternary mixed crystal Nb$_{0.05}$W$_{0.95-x}$Mo$_x$(Se$_{1-x}$S$_x$)$_2$, and they obtained *ZT* ~ 0.63 at 850 K.[47] However, the majority of earlier research on WSe$_2$ concentrated on modifying the *PF* with comparatively large amounts of $\kappa$. Therefore, in order to improve the *TE* efficiencies for WSe$_2$, one must increase the *PF* while simultaneously reducing the $\kappa$ sharply.

In this study, our research endeavors have yielded a significant breakthrough in the domain of TE materials. By employing a co-doping strategy involving the co-doping of Nb and Te into WSe$_2$, we have significantly improved its thermoelectric performance, especially in higher-temperature ranging between 300 K to 850 K. Our findings reveal a substantial increase in the thermoelectric performance, as indicated by enhanced *PF* and simultaneously reduced $\kappa$, the enhancement of the *PF* is linked to enhanced density of state, effective mass ($m_d^*$), improved mobility ($\mu$) and elevated electrical conductivity ($\sigma$) owing to partially replacing Se$^{2-}$ with Te$^{2-}$ and the primary reason for the decrease in $\kappa_L$ can be attributed to the large scattering of phonons due to both size and mass fluctuations among the doping elements and particularly at point defects such as Te$_{Se}$ and Nb$_W$. At last, the achievement of a remarkable maximum figure of merit ZT$_{max}$ ~ 1 at 850 K for the W$_{0.95}$Nb$_{0.05}$Se$_{1.7}$Te$_{0.3}$ sample, underscores the significant potential of our work in advancing thermoelectric applications by co-doping in WSe$_2$ at elevated temperatures.

## 2. Experimental Section

### 2.1. Material Synthesis:

To obtain high-purity tungsten (W), we initially employed $WO_3$. The required quantity of $WO_3$ powder was placed inside a tube furnace. Under a hydrogen atmosphere, the temperature of the tube furnace was gradually increased to 1173 K at a rate of 5 K/min and held at that temperature for 1.5 hours. For the synthesis of different $W_{0.95}Nb_{0.05}Se_{2-y}Te_y$ series materials (where nominal composition y= 0, 0.1, 0.2, 0.3, 0.4, and pure $WSe_2$), precise amounts of high-purity tungsten (W), selenium (Se), niobium (Nb), and tellurium (Te) were individually weighed in ambient air. These weighed components were thoroughly mixed and ground using an agate mortar before being placed into evacuated quartz ampules, which were then sealed. The sealed ampules were inserted into a horizontal furnace. The furnace temperature was subsequently raised to 1173 K at a steady rate of 4 K/min, where it was maintained for an additional 4 hours. Following this, the furnace temperature was reduced to 923 K and held at this temperature for an additional 48 hours before naturally cooling to room temperature. To ensure optimal homogeneity in the composite samples, the resulting output powder underwent further grinding using an agate mortar for a duration of 15 minutes. Subsequently, the fine powder was placed into a graphite die with a diameter of 15 mm and pelletized using spark plasma sintering (SPS) under a pressure of 50 MPa at 1123 K in an argon (Ar) atmosphere.

### 2.2. Structural Characterizations:

Field emission scanning electron microscopy (FE-SEM, Model: SU8020), high-resolution transmission electron microscopy (HR-TEM, Model: JEOL JEM-2010) operating at (200 kV) accelerating voltages, and X-ray diffraction (XRD) using an X-ray diffractometer (Model: Philips X'Pert PRO) with Cu-K radiations (1.54056) were used to conduct both structural and phase analysis of the material. A step size of 0.0263°/min was used to acquire the XRD data for the powder specimen, which covers the angle range of 2θ = 10° to 90°. By combining HR-TEM and energy-dispersive spectroscopy (EDS), samples elemental composition was confirmed. The lattice parameters *a* and *c* were determined using Rietveld refinements using the generalized structural analysis system (GSAS-II). Both the HR-TEM analysis and the Geometric-Phase Analysis (GPA) was performed using the Gatan Digital-Micrograph (GMS-3) system.

## 2.3. Thermoelectric Characterizations:

Using the commercial ZEM-3 (ULVAC-RIKO) equipment, the thermopower ($S$) and electrical conductivity ($\sigma$) were determined in a helium atmosphere across the high temperature ranges from 300 K to 850 K. The accuracy limitations for $S$ and $\sigma$ data are both approximately ±5%. Using the commercial LFA (Netzsch, LFA-457) equipment, the laser flash method was utilized to measure thermal diffusivity ($D$) between 300 K to 850 K. While the equation $\kappa = C_P D d$ was used to measure the total thermal conductivity ($\kappa$), where $C_P$, $D$, and $d$ represent the specific heat, thermal diffusivity, and density, respectively. The accuracy of the measurement of $\kappa$ is about ±7%. The thermal conductivity ($\kappa$) of the hole component ($\kappa_e$) was calculated using the Wiedemann-Franz formula ($\kappa_e = \sigma L T$), where $L$ denotes the Lorentz number. The value of $\kappa_L$ was determined using the equation $\kappa_L = \kappa - \kappa_e$. The Archimedes method was used to calculate the density of all doped specimens in absolute ethanol. The specific heat ($C_P$) was measured using a diamond Differential Scanning Calorimeter (DSC). The relative densities of all doped specimens are 90%. The Hall coefficient ($R_H$) was tested using the physical properties measuring system (PPMS-9) in a reversed magnetic field ($H = 3.0$ T) at 300 K; Hence the accuracy of the measurement of $p$ is about ±20 %. The carrier concentrations ($p$) and carrier mobility ($\mu$) were obtained using the formulae $p = 1/eR_H$ and $\mu = \sigma/ep$, respectively. All statistical thermoelectric property calculations were performed using Origin 2023 (Origin Lab Corporation, Northampton, MA, USA).

## 2.4. Model and Computational Details

In this investigation, electronic structure calculations were conducted using the Density Functional Theory (DFT) method implemented in the Vienna *ab-initio* Simulation Package (VASP).[48,49] The ion-electron interactions employed the Projector Augmented Wave (PAW) pseudo-potential, and electron-electron interactions were described using a plane-wave basis set.[50] Atomic structural relaxation utilized the Generalized Gradient Approximation (GGA) method with the PBE0 model,[51] which is a hybrid functional that incorporates a portion of exact Hartree-Fock exchange into the GGA framework of Perdew–Burke–Ernzerhof (PBE).[52] The plane wave basis was configured with a kinetic energy cutoff of 520 eV, and a Monkhorst-Pack grid of 4x4x1 was employed for Brillouin zone sampling. The initial structure is a unit cell of (001) WSe$_2$ consisting of 3 atoms, with lattice parameters $a = b = 3.28$ Å and $c = 6.48$ Å, taken from experimental values of this work. This unit cell is replicated along the $a$ and $b$ directions to construct a monolayer supercell containing 60 atoms. This supercell is subsequently modified to form the W$_{19}$NbSe$_{34}$Te$_6$ compound, enabling the study of its structural and electronic properties. To ensure adequate separation between repeated images, a 15 Å vacuum was applied between slabs along the (001) direction. The convergence criterion for electronic relaxation was set at $10^{-5}$ eV/cell with a Gaussian smearing and width of 0.02 eV, allowing for a smooth convergence in calculations and avoiding an artificial gap.[53] Internal coordinates were relaxed using conjugate gradient methods until the Hellmann–Feynman forces were less than 0.02 eV/Å. Notably, all atoms in the systems underwent complete relaxation, and no atomic positions were constrained during the optimization process. The Density of States (DOS) were obtained by the tetrahedron method with Blöchl corrections. The transport properties were calculated using the semiclassical Boltzmann transport theory for the supercell of both systems, applying the constant scattering time and rigid band approximations, as implemented in the BoltzTrap code process.[54]

## 3. RESULTS & DISCUSSION

### 3.1. Microstructural Characterizations:

Figure 1(a) exhibits the X-ray diffraction (XRD) spectrum of as-synthesized $WSe_2$ and $W_{0.95}Nb_{0.05}Se_{2-y}Te_y$ (y=0, 0.1, 0.2, 0.3, and 0.4) samples as well as standard PDF card No. 01-087-2418 at 300 K. Except for y=0.4, all XRD peaks for $WSe_2$ and all samples of $W_{0.95}Nb_{0.05}Se_{2-y}Te_y$ (0≤y≤0.03) are in full accordance with standard PDF card, with no additional phase impurity being found. There are additional peaks for y=0.4, though, indicating by blue star, which show the presence of $NbTe_2$ phase in $W_{0.95}Nb_{0.05}Se_{2-y}Te_y$. As displayed in Figure 1b, which shows the enlarged XRD pattern from 2θ=10º to 40º for each specimen, all the peaks of XRD patterns shift toward lower 2θ angles after doping of Nb on W site and Te on Se site, and the peak shift increases as we increase doping concentration. In order to better estimate the lattice parameters for $W_{0.95}Nb_{0.05}Se_{2-y}Te_y$ samples, Rietveld refinements on XRD data were performed (Figure S4). Figure 1c shows the lattice parameters *a* and *c* that were derived by Rietveld refinements. We can conclude from Figure 1c that both of the lattice parameter *a* and *c* increase consistently as we increase the doping concentration of Te (for instance *a* increases from 3.2820 Å for $WSe_2$ to 3.2881 Å for y=0, 3.2969 Å for y=0.1, 3.3054 Å for y=0.2, 3.3178 Å for y=0.3, and 3.3296 Å for y=0.4; similarly, c increases from 12.96 Å for $WSe_2$ to 13.005 Å for y=0, 13.092 Å for y=0.1, 13.1375 Å for y=0.2, 13.1937 Å for y=0.3, and 13.2486 Å for y=0.4). The reason for this change in crystal lattice is actually due to the slightly large ionic radii of $Nb^{3+}$ ($r_i$=0.72Å)[55] than $W^{4+}$ ($r_i$=0.66Å)[55] and also high ionic radii of $Te^{2-}$ ($r_i$=2.21Å)[55] than $Se^{2-}$ ($r_i$=1.98 Å)[55]. Thus, the current finding suggests that Nb is substituted for W and Te for Se, leading to unit cell expansion for $WSe_2$ lattice. Moreover,

Rietveld refinement was further performed (as shown in Figure S4) for the all samples with different *y* content to estimate the Se vacancies ($V_{Se}^{\bullet\bullet}$) within the samples. The obtained Se vacancies ($V_{Se}^{\bullet\bullet}$) are shown in Figure S5 (see Supporting Information) as a function of Te content *y*. Thus, we can conclude from Figure S5 and Table S1 that Se vacancies decreases from $1.95\times10^{20}$ cm$^{-3}$ to $1.12\times10^{20}$ cm$^{-3}$, $1.19\times10^{20}$ cm$^{-3}$, and $0.63\times10^{20}$ cm$^{-3}$ with increasing Te doping content *y* from 0 to 0.1, 0.2, and 0.3, then Se vacancies is further increased to $1.20\times10^{20}$ cm$^{-3}$ for *y*= 0.4 due to presence of another phase impurity ($NbTe_2$) which is confirmed by XRD. This reduction of Se vacancies ($V_{Se}^{\bullet\bullet}$) (for 0≤*y*≤0.3) is primarily caused by formation energy of Se vacancies, this phenomenon is also attributed to the increased ionic radii of Te as compared to Se.

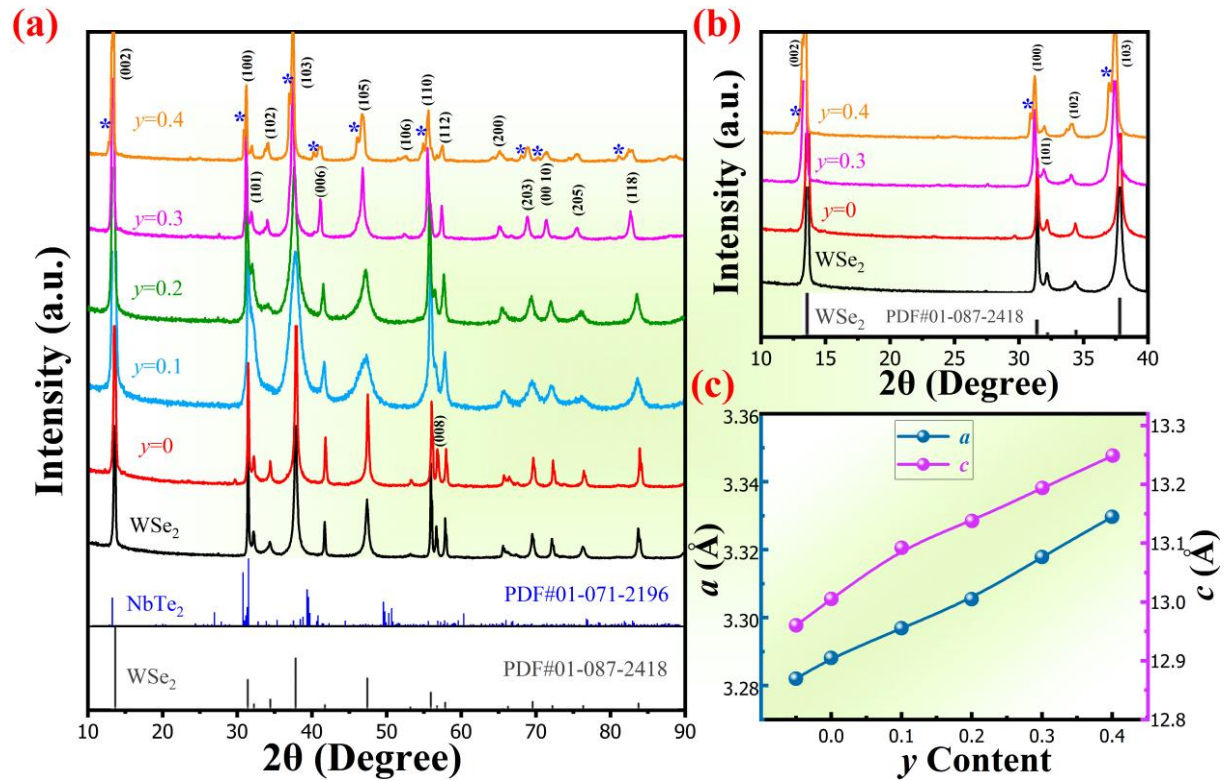

**Figure 1.** (a) XRD patterns for the co-doped $W_{0.95}Nb_{0.05}Se_{2-y}Te_y$ system, where y=0, 0.1, 0.2, 0.3, & 0.4 and blue stars indicate the existence of Te; (b) magnified peaks from XRD for the specimens of $WSe_2$, *y*=0, *y*=0.3, & *y*=0.4, and (c) the lattice parameter values a & c for each co-doped system.

In order to gain additional insight into the TE performance of $W_{0.95}Nb_{0.05}Se_{2-y}Te_y$, its microscopic characteristics were further examined by using FE-SEM as well as HR-TEM. Figure S1 (see Supporting Information) exhibits the morphological characteristics of the fractured surface of $W_{0.95}Nb_{0.05}Se_{2-y}Te_y$ (0≤$y$≤0.4) bulk samples soon after SPS process, as analyzed by FE-SEM. According to the FE-SEM findings, each sample has a distinct multi-layered hexagonal structure including well-defined, sharply edged nano-size grain sheets. Furthermore, energy dispersive spectroscopy (EDS) has been used to investigate different elemental compositions. EDS mapping (see supporting Information section, Figure S3 (a), (b), (c), (d), (e) and (f)) demonstrates that Nb and Te is perfectly doped and uniformly substituted in the specimen's surface. Additionally, EDS mapping indicates that each atom exactly matches the formula ratio.

In order to learn more about the microstructures of a typical co-doped specimen with $y$=0.3, the transmission electron microscopy with high resolution (HR-TEM) data were studied. Figures 2a and 2b demonstrate that all grains have layered nano-sheet with prominent grain

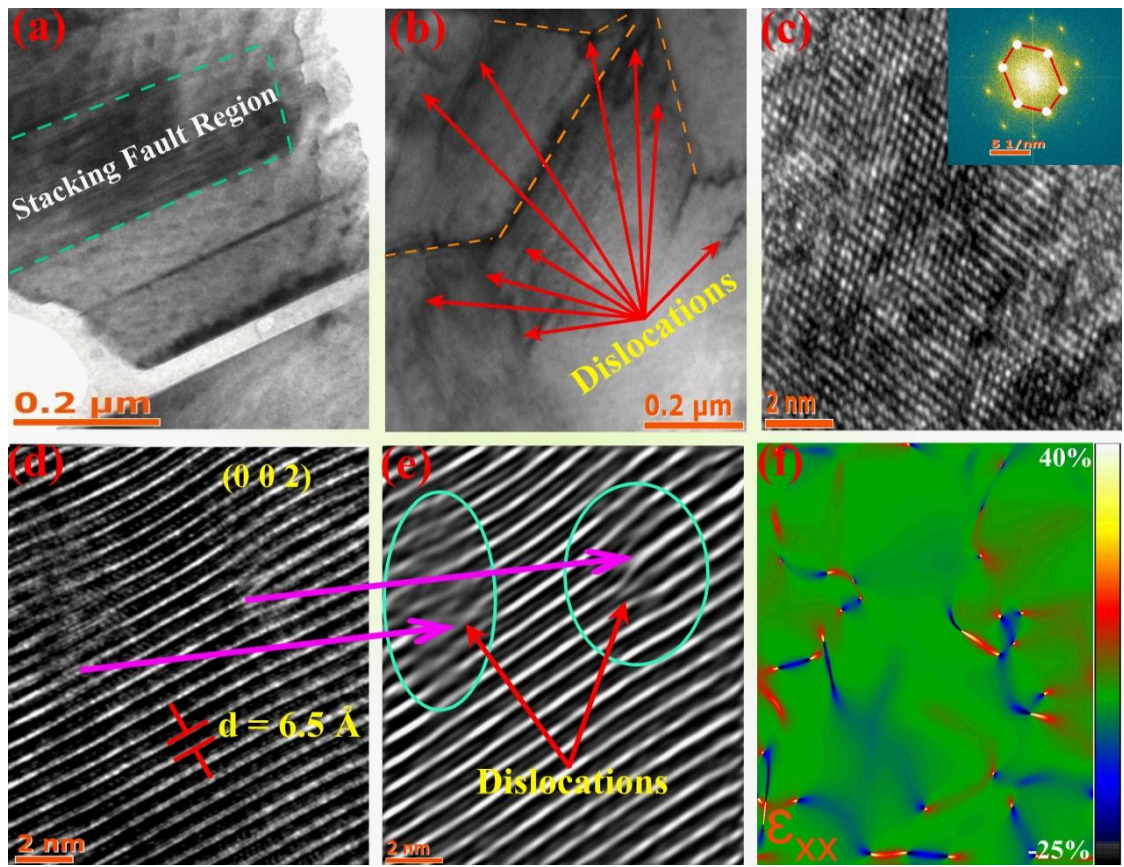

**Figure 2**. (a & b) displays low-resolution TEM images that clearly show grain boundaries with multiple dislocations, (c) reveals high-resolution TEM images of a typical sample at $y$=0.3, with an inset of the Fast Fourier Transform (FFT) image from the same site that also demonstrates hexagonal structure, (d) TEM High-resolution images, which reveals the various dislocation regions, (e) An inverse Fast Fourier transform (IFFT) image of image d, which also shows the correlates dislocation areas given in image d, and (f) Corresponding strain mapping

($\varepsilon_{xx}$) of image (d) along the xx direction, indicating the dislocation region; the scale bar displays the strain changes (color changes from blue to yellow); positive data indicates the strains in tensile analysis, and negative data indicates strains in compression analysis.

boundaries and multiple dislocation regions. the Figure 2c shows the high-resolution micrograph across the site of $W_{0.95}Nb_{0.05}Se_{1.7}Te_{0.3}$, the inset shown in Figure 2c displays the IFFT image which also indicates the hexagonal structure of the substance, which seems in accordance with XRD and FE-SEM investigation. Meanwhile, the high-resolution picture (In Figure 2d) clearly demonstrates the large number of dislocations regions. An inverse Fast Fourier transform (IFFT) image of the dislocation zones of Figure 2d is given in 2e. A comprehensive analysis of the HR-TEM images (Figure 2d) reveals the presence of a single phase; no additional impurity phases are observed. Using crystal data, the inter planer crystal spacing (d=6.5 Å) has been calculated (see Figure 2d). The reciprocal length and angles correspond to the simulated results. In order to identify the geometric strain regions resulting from mechanical deformations, the strain analysis (SA) and Geometric Phase Analysis (GPA) of the HR-TEM images were also examined. The strain map produced by GPA (Figure 2f) clearly distinguishes the areas of dislocations under tensile along with compression strains. Figure 2f depicts the strain field mapping ($\varepsilon_{xx}$) of Figure 2d across the xx directions, where negative quantities represent compression strains while positive data represent tensile strains.

### 3.2. Thermoelectric Characterizations:
### 3.2.1. Electronic Transport Parameters:

The temperature dependent electrical conductivity for all doped samples of $W_{0.95}Nb_{0.05}Se_{2-y}Te_y$ (*y*=0,0.1,0.2,0.3, and 0.4) is shown in Figure 3a. Electrical conductivity decreases with increasing Te doping content, as seen in Figure 3a. The electrical conductivity, for instance, $\sigma$ is 174 $\Omega^{-1}cm^{-1}$, 110 $\Omega^{-1}cm^{-1}$, 104 $\Omega^{-1}cm^{-1}$, 100 $\Omega^{-1}cm^{-1}$, and 2.5 $\Omega^{-1}cm^{-1}$, for the specimens with *y* = 0, 0.1, 0.2, 0.3, and 0.4 at room temperature, respectively. One can see in Figure 3a, the value of $\sigma$ for single doped (only Nb doped) sample is decreasing with increasing temperature, indicating the degenerate behavior, while for all co-doped samples $\sigma$ rises with temperature as its highest temperature T=850 K is reached, demonstrating that all co-doped samples belong to partially degenerate semiconductors. This partially degenerate behavior of all co-doped samples is also confirmed from the Figure 3b, where the *p* (usually *p* is in the range of $10^{18}$ cm$^{-1}$ for partially degenerate semiconductors) for the co-doped samples is much lower as compared to Nb doped (*y*=0) sample. In order to better understand the changing behavior of the $\sigma$, the Hall coefficient $R_H$ was measured for all of the bulk specimens of $W_{0.95}Nb_{0.05}Se_{2-y}Te_y$ at room temperature. The obtained $R_H$ and derived hole concentration $p$ and hole mobility $\mu$ are shown in Figure 3b & Table S2 as functions of Te content $y$. As the Te doping content rises from $y=0$ to $y=0.04$, the hole concentration $p$ decreases monotonically from $170\times10^{18}$ cm$^{-3}$ to $1.19\times10^{18}$ cm$^{-3}$. In contrast, $\mu$ rises from 6.14 cm$^2$ V$^{-1}$ s$^{-1}$ to 44.83 cm$^2$ V$^{-1}$ s$^{-1}$ for $y=0$ to $y=0.3$; after that it decreases to 7.77 cm$^2$ V$^{-1}$ s$^{-1}$ for $y=0.4$ (this decreased $\mu$ for $y=0.4$ is caused by large scattering of carriers with impurity phase). Thus, the decreased $\sigma$ can be attributed to reduction of hole concentration $p$. In contrast to our earlier research, which shown that only Nb doping in WSe$_2$ increases $p$, the current work demonstrates that $p$ decreases with increasing Te content. This decrease in $p$ could also be explained by behavior of Te as a donor atom in WSe$_2$.[56]; while the increased $\mu$ could be ascribed to weakened carrier scattering by reduced Se vacancies $V_{Se}^{\bullet\bullet}$ (as shown in Figure S5). Besides, the simultaneous substitution of different dopant elements as co-doping can lead to synergistic effects, i.e., it produces unique electronic and crystallographic modifications, which enhances the electronic density of state (E-DOS) (for more details, see DFT Observation section); This increased E-DOS should also contribute to heightened *μ*, augmenting the probability of efficient carrier movement within the material.[57] these enhanced states may create a more favorable environment for carrier transport, contributing to enhancement of mobility and thermopower which result into high *PF*.

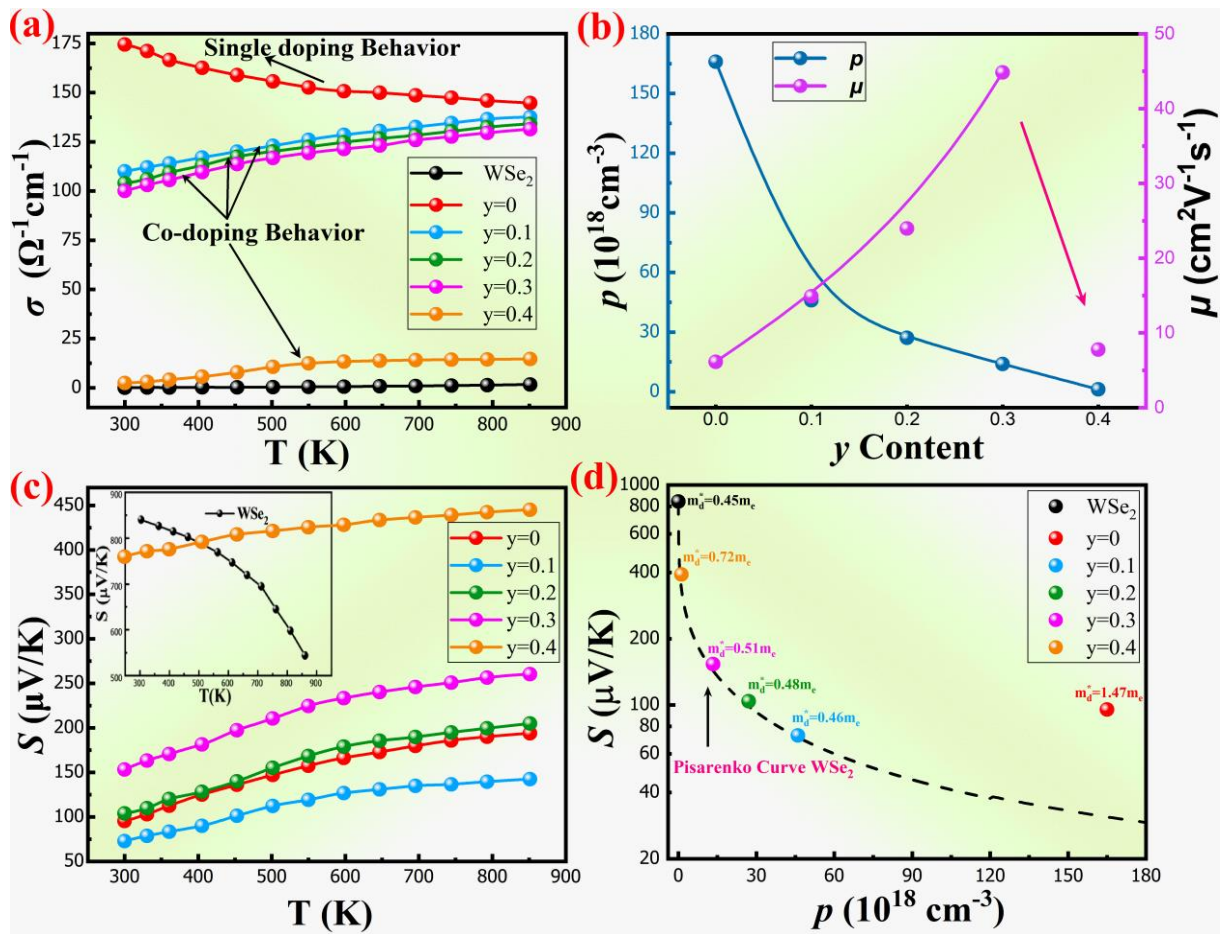

**Figure 3**. (a) Temperature dependent electric conductivity ($\sigma$) for all co-doped samples of $W_{0.95}Nb_{0.05}Se_{2-y}Te_y$ along with WSe$_2$ sample, (b) variation of carrier mobility ($\mu$) and carrier concentrations ($p$) at 300 K along $y$ contents, (c) temperature-dependent thermopower ($S$) for all co-doped systems, with the inset representing the *S* for WSe$_2$, and (d) variation of *S* with hole carrier concentration (*p*) at 300 K (Pisarenko curve) therein, the effective mass ($m_d^*$) of each sample is also given.

Figure 3c shows the thermopower (*S*) coefficient for each W$_{0.95}$Nb$_{0.05}$Se$_{2-y}$Te$_y$ sample. All of the *S* values for doped and pristine system are positive, indicating that this material is p-type semiconducting. The absolute value of *S* for all doped systems rises with temperature as its highest temperature T=850 K is reached, in contrast to the electrical conductivity, demonstrating that all co-doped systems belong to partially degenerate semiconductors. On the other hand, pure WSe$_2$ has *S* coefficient that decreases with increasing temperature, indicating the typical non-degenerative behavior. The highest *S* coefficient for the doped system is 392 μ V K$^{-1}$ at 300 K and 445 μ V K$^{-1}$ at 850 K for *y*=0.4. One can see from Figure 4c that basically as Te doping increases *S* increase (for instance, *S* at 300 K rises from 72 μ V K$^{-1}$ to 101 μ V K$^{-1}$, 153 μ V K$^{-1}$, and 392 μ V K$^{-1}$ as *y* increases from 0.1 to 0.2, 0.3, and 0.4, respectively). The increase of *S* for the doped specimens can basically be explained by decreased hole concentration. According to Mott formula, *S* has the relation with *p*, as follows:[58,59]

$$S = \frac{\pi^2 k_B^2 T}{3q} \left( \frac{\partial \ln(\sigma(E))}{\partial E} \right)_{E=E_f} \quad (1)$$

$$S = \frac{\pi^2 k_B^2 T}{3q} \left( \frac{1}{p} \frac{\partial p(E)}{\partial E} + \frac{1}{\mu} \frac{\partial \mu(E)}{\partial E} \right)_{E=E_f} \quad (2)$$

The Mahan-Sofo theory states that the formula (2) can be simplified as follows:[60]

$$S = \frac{8\pi^2 k_B^2 T}{3qh^2} m_d^* \left( \frac{\pi}{3p} \right)^{2/3} \quad (3)$$

where $m_d^*$ stands for an E-DOS effective mass, *p(E)* stands for the Hall carrier density, *μ(E)* for the Hall carrier mobility, $E_f$ for the Fermi energy, $k_B$ for the Boltzmann coefficient, and for the charge carrier. Thus, equation 1 and 2 show that the *S* is inversely proportional to *p*, which could explain why *S* increases with increasing of Te content *y* due to decreasing trend *in p*. However, the fact that *S* of the doped sample with *y*=0 is greater than that of the doped sample with *y*=0.1 seems contradictory to formula 2. This abnormal phenomenon implies that some other mechanism works in this single doped (*y*=0) sample. Since *S* depends on multiple variables, and whether *S* is influenced by other factors (such as E-DOS or $m_d^*$) would be further explored. To further understand the mechanism behind the increase in the behavior of *S* for *y*=0 and also for increasing Te concentration in co-doped systems, we calculated the E-DOS $m_d^*$ by the following formulae.[61,62]

$$m_d^* = \frac{h^2}{2k_B T}\left(\frac{p}{4\pi F_{1/2}(\eta)}\right)^{2/3} \tag{4}$$

$$S = \frac{k_B}{q}\left[\frac{(\lambda+2)F_{(\lambda+1)}(\eta)}{(\lambda+1)F_\lambda(\eta)} - \eta\right] \tag{5}$$

$$F_j(\eta) = \int_0^\infty \frac{x^j}{1-e^{x-\eta}}dx \tag{6}$$

where $\eta$ denotes the reduced Fermi level, which can be defined as $\eta = E_f/k_B T$, $F_j(\eta)$ indicates the jth order Fermi Integral, and *h* stand for the Planks constant. In this case, we suppose that all of the co-doped samples exhibit prominent acoustic phonon scattering, with λ=0. On the basis of this assumption, the Pisarenko curve for WSe$_2$ can then be calculated, as seen by the dashed line in Figure 3d. Thus, Figure 3d shows that *S* rises by 214% (from 30.16 μ V K$^{-1}$ to 95.3 μV K$^{-1}$) for *y*=0, 2% (from 71.3 μ V K$^{-1}$ to 72.7 μV K$^{-1}$) for *y*=0.1, 6% (from 98.42 μ V K$^{-1}$ to 104.04 μ V K$^{-1}$) for *y*=0.2, 9.3% (from 140.2 μ V K$^{-1}$ to 153.22 μ V K$^{-1}$) for y=0.3, and 17.8% (from 333.2 μ V K$^{-1}$ to 392.4 μ V K$^{-1}$) for *y*=0.4 as in comparison to *S* of pure WSe$_2$, respectively. One may compute $m_d^*$ for all the samples through the experimental data of *p* and *S* at 300 K and the assumption that λ=0 (i.e., acoustic phonon scattering behavior is dominant) for the all co-doped systems, as shown in Figure 3d. As revealed in Figure 3d, the $m_d^*$ rises from 0.45m$_e$ for WSe$_2$ to 0.46m$_e$ for *y*=0.1, 0.48m$_e$ for *y*=0.2, 0.51m$_e$ for *y*=0.3, 0.72m$_e$ for *y*=0.4, and 1.47m$_e$ for *y*=0. This large enhancement of $m_d^*$ in *y*=0 shows that *S* of *y*=0 is greater than that of the doped sample with *y*=0.1. According to formula 2 and 3, this obvious increase in $m_d^*$ and decrease in *p* is responsible for the increase in *S* with increasing *y* and advantageous to the resulting increase in the *PF* and *ZT* measurements. This trend is also observed in DFT calculations and added in Supporting Information Section.

### 3.2.2. DFT Investigation on Structural and Electronic Transport Parameters

In WSe$_2$, W atoms and Se atoms occupy the sub lattices of the hexagonal sheet, with Se atoms on the lower layer positioned indirectly beneath those on the upper layer, corresponding to the P63/mmc space group. Calculations reveal a W-W bond length of 3.31 Å and a W-Se bond length of 2.54 Å. As illustrated in Figure 4c, the lattice parameters exhibit in-plane (a=b=3.28

Å) and perpendicular (c=6.48 Å) dimensions, aligning with experimental measurements in Figure 1c. Figure 4c illustrates the unit cell of WSe$_2$, which was geometrically relaxed using experimental lattice parameters. Subsequently, a monolayer cell was generated and duplicated, as detailed in the Methods section. The separation between monolayers in the supercell is 15 Å, as depicted in the side view, with each layer consisting of 60 atoms. The electronic band structure and density of states (DOS) calculated for WSe$_2$, as depicted in Figure 4a and 4b, respectively, reveal its non-magnetic semiconducting nature with an indirect band gap of 1.29 eV. These findings agree well with reported literature data and our own experimental results.[63]

We investigated the electronic properties of WSe$_2$ doped with 5% Nb and 15% Te, mirroring experimental conditions. The unit cell, replicated 4x3x1, underwent random substitution of 15%

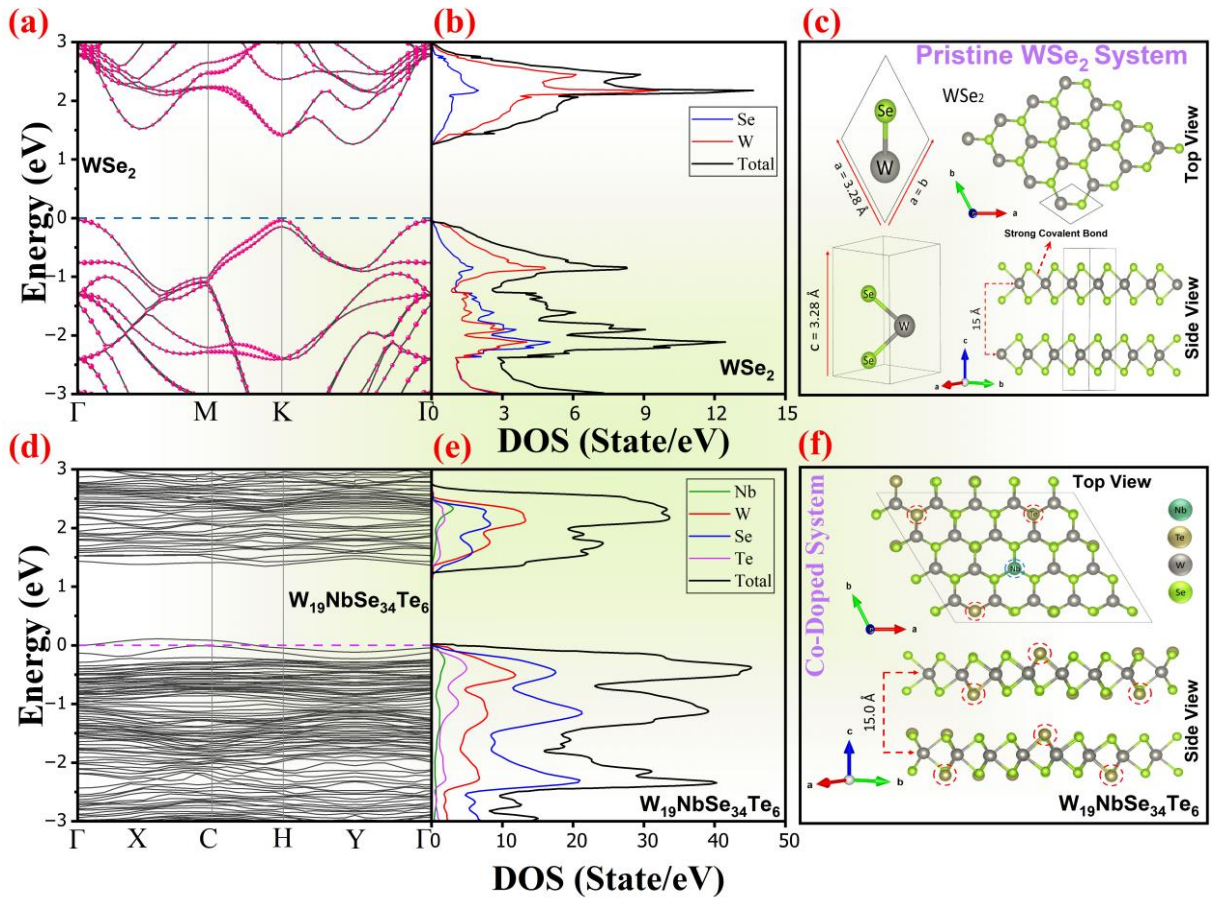

**Figure 4.** (a-b) Band structure and Density of States of the optimized unit cell of WSe$_2$, (c) showcasing excellent agreement with the lattice parameter of a=3.28 Å (in plane) and c=6.48 Å (perpendicular), in alignment with experimental measurements as shown in Figure 1c. (d-e) Band Structure and Density of States for -4.5 to 2.5 eV with a focus on -2 to 2 eV, revealing electronic band structures for the path Γ, X, C, H, Y, Γ due to the atomic arrangement. While remaining semiconducting, Nb-Te-doped WSe$_2$ displays additional electronic states, reducing

the energy gap to 1.22 eV compared to pristine WSe$_2$ of 1.28 eV. (f) Atomic structure of 5% of Nb and 15% of Te doped WSe$_2$ obtaining NbW$_{19}$Se$_{34}$Te$_6$, evenly distributed in the top and bottom monoatomic layers.

Se atoms with Te atoms at 7.5% each in the top and bottom monoatomic layers. Only a single W atom that corresponds to 5% of W is substituted by Nb for each monolayer resulting in a composition of W$_{0.95}$Nb$_{0.05}$Se$_{1.7}$Te$_{0.3}$, as depicted in Figure 4f. After the energy optimization process, the lattice parameters were determined (a=3.31 Å, c=6.65 Å), in fair agreement with experimental data (Figure 1c). Figure 4d-e illustrates the density of states across an energy range of -4.5 to 2.5 eV, with a zoom from -2 to 2 eV, providing a clearer view of the electronic band structures for the W$_{19}$NbSe$_{34}$Te$_6$ system choosing a path of Γ, X, C, H, Y, Γ due to the geometry of the compound. Despite remaining semiconducting, Nb-Te-doped WSe$_2$ exhibits additional electronic states within the gap region, reducing the energy gap to 1.22 eV compared to pristine WSe$_2$. The reduction in the energy gap is found to be directly proportional to an elevation in the density of states, a trend consistent with experimental findings indicating an increased reservoir of energy states available to carriers. This surplus of states contributes to heightened carrier mobility, augmenting the probability of efficient carrier movement within the material.[57] The band structure, intricately tied to the electronic density of states, plays a pivotal role in co-doping of WSe$_2$ with Te and Nb. The co-doped TE system W$_{19}$NbSe$_{34}$Te$_6$ shows favorable band structures as compared to pristine WSe$_2$, particularly characterized by a high density of states near the band edges, fostering more efficient carrier transport. A higher density of states proximate to the conduction or valence band edges positively influences carrier mobility and thermopower, whereas the presence of tails in the band structure has a negative impact. A more abrupt change in the band structure signifies superior transport properties. Thus, Present experimental findings are well synchronized with computational results which demonstrate that co-doping causes the elevated E-DOS (or enhanced $m_d^*$) and leading to enhancement of *S*.

### 3.2.3. Thermal Transport Parameters

To achieve the remarkable TE characteristics of the co-doped W$_{0.95}$Nb$_{0.05}$Se$_{2-y}$Te$_y$ systems in our work, the especially low κ also played a significant role. Figure 5a represents the behavior of the total κ for all co-doped specimens of W$_{0.95}$Nb$_{0.05}$Se$_{2-y}$Te$_y$. It can be seen that the κ drops as the Te content *y* increases. In contrast to pristine WSe$_2$, all co-doped samples have possessed significantly lower *κ* in the whole temperature range investigated. For instance, at 300 K, as *y* enhances from 0 to 0.1, 0.2, 0.3, and 0.4, then κ decreases from 1.83 W m$^{-1}$ K$^{-1}$ to 1.62 W m$^{-1}$ K$^{-1}$, 1.48 W m$^{-1}$ K$^{-1}$, 1.29 W m$^{-1}$ K$^{-1}$, and 1.17 W m$^{-1}$ K$^{-1}$, respectively; while pristine WSe$_2$ possesses κ = 3.35 W m$^{-1}$ K$^{-1}$ at 300 K, which indicates 65% decrease in total κ. The total κ is the sum of the effects of κ$_L$ and κ$_e$, denoted as κ = κ$_e$ + κ$_L$. The value of κ$_e$ can

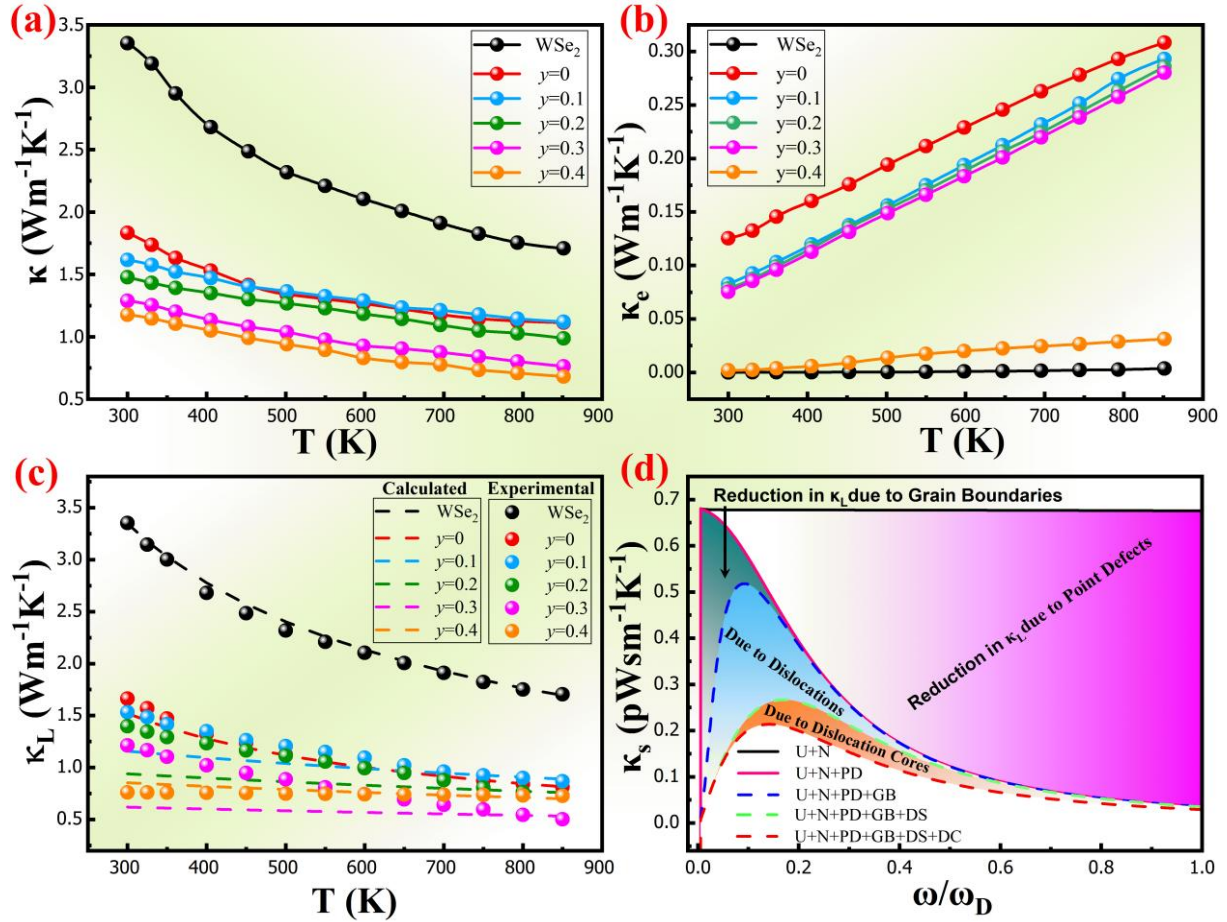

**Figure 5** (a) Temperature dependent total thermal conductivity (κ) for all co-doped W$_{0.95}$Nb$_{0.05}$Se$_{2-y}$Te$_y$ systems compared to WSe$_2$, (b) variation of electronic thermal conductivity (κ$_e$) with temperature, (c) the corresponding comparison across theoretical lattice thermal conductivity (in lines) by applying Debye-Callaway model and data from experiments (in symbols), and (d) derived spectral thermal conductivity (κ$_s$) based on experimental analysis with phonon frequency for the co-doped sample of *y*=0.3.

be derived by using Weidman-Franz's law (κ$_e$= σ*LT*), where *L* represents the Lorentz number. For all co-doped specimens, the magnitude of *L* can also be determined using the following relationship:[64]

$$L = \left\{1.5 + exp\left(-\frac{|S|}{116}\right)\right\} \times 10^{-8} \text{ W}\Omega\text{K}^{-2} \quad (7)$$

Figure 5b shows the experimental measurement for $\kappa_e$, which shows that $\kappa_e$ increases monotonically with temperature and correspondingly $\kappa_e$ decreases when Te concentration increases as a result of decreased $\sigma$. Subsequently, the value of $\kappa_L$ can potentially be calculated using the equation $\kappa_L = \kappa - \kappa_e$. Figure 5c demonstrates that when Te content rises, $\kappa_L$ reduces. At 300 K, for instance, the related values of $\kappa_L$ reduce from $\kappa_L$=3.35 W m$^{-1}$ K$^{-1}$ for WSe$_2$ to 1.66 W m$^{-1}$ K$^{-1}$ for $y$=0, 1.53 W m$^{-1}$ K$^{-1}$ for $y$=0.1, 1.39 W m$^{-1}$ K$^{-1}$ $y$=0.2, 1.21 W m$^{-1}$ K$^{-1}$ $y$=0.3, and 0.76 W m$^{-1}$ K$^{-1}$ for $y$=0.4, indicating that Nb and Te co-doping causes significant and continuous reduction of lattice thermal conductivity. However, at high temperature ranges (above 650 K) the $\kappa_L$ of $y$=0.4 is larger than $y$=0.3 because of the presence of impurity phase, indicating that large content of $y$ is not suitable for this W$_{0.95}$Nb$_{0.05}$Se$_{2-y}$Te$_y$ systems.

In order to gain a deep understanding why the experimental measurements of $\kappa_L$ decreases significantly for all co-doped W$_{0.95}$Nb$_{0.05}$Se$_{2-y}$Te$_y$ systems, the Debye-Callaway Model (Formula 8; for additional details for the calculations performed by the Debye-Callaway Model, see the corresponding subsection in Supporting Information) was implemented for fitting the experimental findings of $\kappa_L$ in the whole temperature range for all co-doped samples. The Debye-Callaway Model is defined as following:[65]

$$\kappa_L = \frac{2\pi k_B}{h} T^3 \int_0^{\frac{\theta_D}{T}} \frac{x^4 e^\beta}{\tau^{-1}(e^\beta - 1)^2} d\beta \qquad (8)$$

Where $T$ is the temperature, $k_B$ is the Boltzmann constant, $\tau$ is referred as the phonon scattering relaxation time, $h$ stands for the Planks constant, and $\beta$ shows the absolute phonon frequency, defined as $\beta = h\omega/2\pi k_B T$. In this theoretical calculation approach, we analyzed the phonon-phonon Umklapp scattering (U-scattering) and normal scattering (N-scattering), point defect scattering (PD), grain boundary scattering (GB), dislocation scattering (DS), and dislocation core scattering (DC). The overall phonon scattering relaxation time can potentially be expressed using the Mattheissen's relation.[66]

$$\tau^{-1} = \tau_U^{-1} + \tau_N^{-1} + \tau_{GB}^{-1} + \tau_{DC}^{-1} + \tau_{DS}^{-1} + \tau_{PD}^{-1} \qquad (9)$$

where $\tau_N$ stands for the relaxation time due to N-scattering, $\tau_U$ stands for the relaxation time due to Umklapp U-scattering, $\tau_{GB}$ stands for the relaxation time due to grain boundaries (GB), $\tau_{DS}$ stands for the relaxation time due to dislocation strain scattering, $\tau_{DC}$ stands for the relaxation time due to dislocation core scattering, and $\tau_{PD}$ stands for the relaxation time due to point defect scattering (the more information of these variables is provided in Supporting Information). Figure 5c shows that over high temperature ranges, theoretically calculated lattice thermal conductivity (shown in lines) and experimentally observed $\kappa_L$ (shown in symbols) of $W_{0.95}Nb_{0.05}Se_{2-y}Te_y$ are in good agreement. To further confirm the consequences of each scattering mechanism on $\kappa_L$, the spectral lattice thermal conductivity ($\kappa_s$) was calculated for $y=0.3$, as shown in Figure 5d. one can notice, each scattering mechanism obviously contributes to lowering the $\kappa_L$, with point defect scattering plays a dominant role as due to co-doping of Nb and Te in $WSe_2$ (Figure 5d).

### 3.2.4. Thermoelectric Power Factor and Figure-of-Merit

The temperature dependent of *PF* for each sample of this co-doped system is shown in Figure 6a. In the present case, the *PF* has significantly enhanced as a consequence of improved $\sigma$ and $\mu$ due to the Te substitution in $WSe_2$. Therefore, the large $PF \approx 8.91$ μ W cm$^{-1}$ K$^{-2}$ at 850 K is achieved for $y=0.3$. Of course, an elevated *PF* result in this co-doped system also indicates improved TE performance. The temperature dependence of figure of merit *ZT* is shown in Figure 6c for each of W$_{0.95}$Nb$_{0.05}$Se$_{2-y}$Te$_y$ specimen. The *ZT* for all co-doped specimens is clearly higher than the results for WSe$_2$ across the entire temperature ranges. However, the *ZT* value of all doped specimens increases as *y* increases until *y*=0.3, after that it decreases for *y*=0.4. As a consequence, a doped specimen with *y*=0.3 achieves the high *ZT$_{max}$*, ~ 1, which is ~1.4-times greater than *y*=0 and ~30 times much higher than that of WSe$_2$. This achievement of *ZT$_{max}$* is the highest value ever reported in the p-type WSe$_2$ TE materials. Additionally, Figure 6b displays the average *ZT$_{ave}$*, values for all incorporated samples, which are determined by integrating the *ZT* curve area within the *ZT* curves (in temperature range from T$_L$=300 K to T$_h$= 850 K) using equation (10):[67,68]

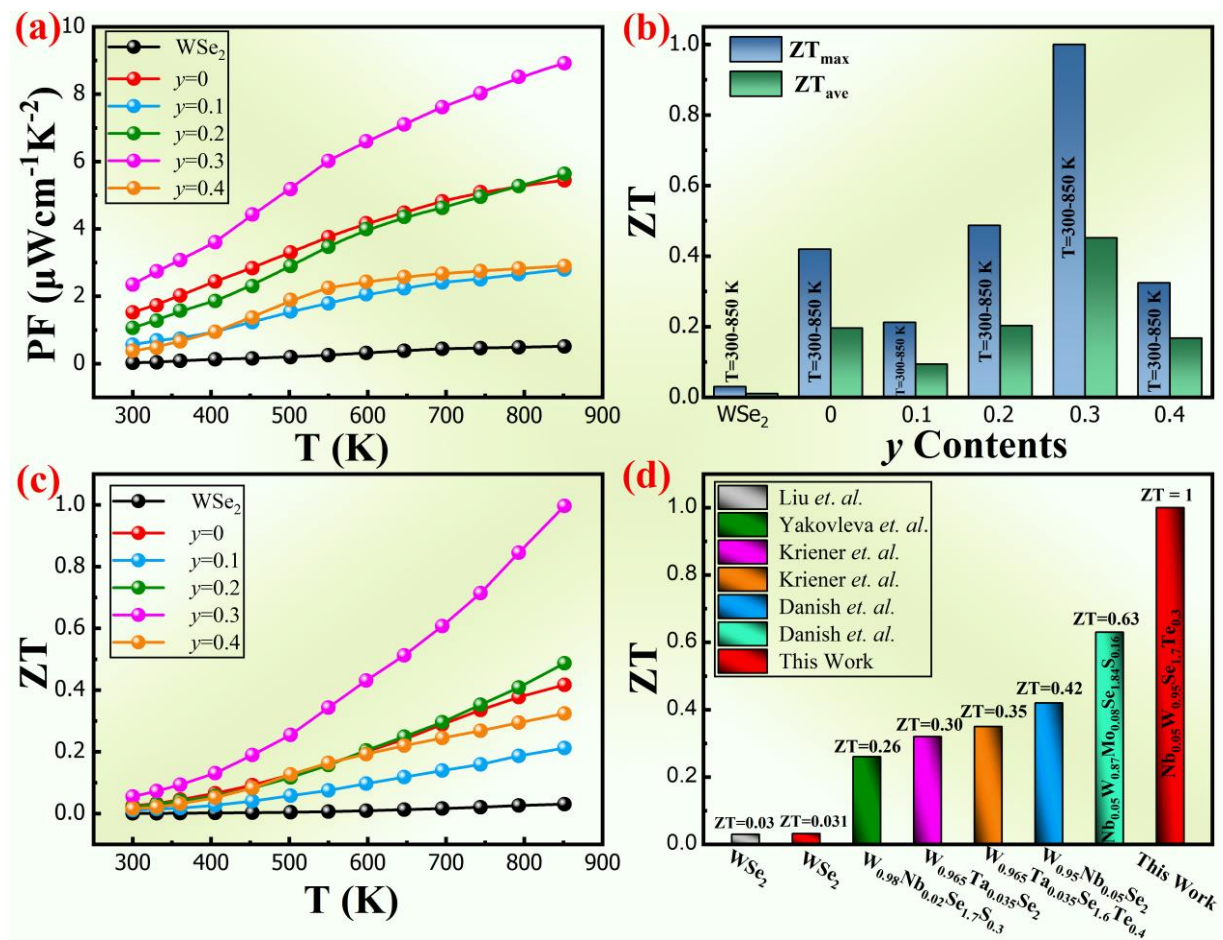

**Figure 6.** (a) Temperature dependence of power factor (*PF*) for all co-doped specimens as compared to WSe$_2$, (b) *ZT$_{max}$* & *ZT$_{ave}$* for all samples at temperature range (300K to 850K), (c) temperature dependence of *ZT* data for all co-doped W$_{0.95}$Nb$_{0.05}$Se$_{2-y}$Te$_y$ samples in comparasion with pristine WSe$_2$, and (d) the maximum *ZT* obtained in the current study is compared with those from previously published studies.

$$ZT_{ave} = \frac{1}{T_h - T_L} \int_{T_L}^{T_h} ZT \, dT \tag{10}$$

Figure 6d compares the maximum ZT$_{max}$, acquired here with that reported in previous research, which demonstrates that *ZT$_{max}$*, attained in the current study is the highest among all the published work. Obviously, this outstanding thermoelectric performance is definitely the result of a significant improvement in *PF* as well as a significant decrease in the lattice thermal conductivity for the co-doped specimens.

## 4. CONCLUSION

In conclusion, we have developed co-doped W$_{0.95}$Nb$_{0.05}$Se$_{2-y}$Te$_y$ system successfully and the doping effects of Nb and Te in WSe$_2$ on its microstructure, electrical, and thermal transport characteristics were studied thoroughly by using experimental and computational approach. The results demonstrate that this co-doping in WSe$_2$ can provide significant increase in the *PF* and large reductions in κ$_L$ simultaneously. The 17-fold enhanced *PF* is a consequence of improved *σ* and *μ* due to the synergistic effect of co-doping of Nb and Te in WSe$_2$. Our results show that κ$_L$ of the co-doped systems can be reduced to 0.48 W m K$^{-1}$ (for *y* = 0.3) at 850 K, which is 72% lower than that of WSe$_2$. Our study based on experimental observations shows that the low κ$_L$ of co-doped system derives mostly from intensive phonon scattering from point defects and strained domains due to high size and mass fluctuations among W and Nb as well as Se and Te. As a result, a remarkable high TE figure of merit $ZT_{max} \sim 1$ at 850 K is obtained for co-doped $W_{0.95}Nb_{0.05}Se_{2-y}Te_y$ system with $y=0.3$, which is revealing a ~ 30-fold increase than that of $WSe_2$, demonstrating that co-doping of Nb and Te in $WSe_2$ is an efficient way to enhance the TE performance of $WSe_2$.

## ACKNOWLEDGEMENTS

This study was financially supported by the National Natural Science Foundation of China through the research funding numbers 12174393, 11674322, 51672278, and 51972307. We would also like to express our gratitude for the assistance provided by the Anhui Provincial Natural Science Foundation through the research funding number 2008085MA18 and the Special Foundation of President of (HFIPS) through the grant number YZJJ202102. We also acknowledge support from the European Union Horizon 2020 research and innovation program under grant agreement no. 857470, from the European Regional Development Fund via the Foundation for Polish Science International Research Agenda PLUS program grant No. MAB PLUS/ 2018/8. We acknowledge the computational resources provided by the High-Performance Cluster at the National Centre for Nuclear Research.

## CONFLICT OF INTEREST

The all authors declare no conflict of interest.

## FUNDING SOURCES

1. ***National Natural Science Foundation of China:*** Grant No. 12174393, 11674322, 51672278 and 51972307.
2. ***Anhui Provincial Natural Science Foundation:*** Grant No. 2008085MA18.
3. ***Special Foundation of President of Hefei Institutes of Physical Science, Chinese Academy of Sciences (HFIPS):*** Grant No. YZJJ-2021-02, and YZJJ-GGZX-2022-01.
4. ***European Union Horizon 2020 research and innovation program:*** Grant No. 857470.
5. ***European Regional Development Fund via the Foundation for Polish Science International Research Agenda PLUS program:*** Grant No. MAB PLUS/ 2018/8.